# Positive versus negative resistance response to hydrogenation in palladium and its alloys.

S. S. Das, G. Kopnov and A. Gerber

Raymond and Beverly Sackler Faculty of Exact Sciences,

School of Physics and Astronomy, Tel Aviv University,

Ramat Aviv, 69978 Tel Aviv, Israel

Resistive solid state sensors are widely used in multiple applications, including molecular and gas detection. Absorption or intercalation of the target species varies the lattice parameters and an effective thickness of thin films, which is usually neglected in analyzing their transport properties in general and the sensor response in particular. Here, we explore the case of palladium-based thin films absorbing hydrogen and demonstrate that expansion of thickness is an important mechanism determining the magnitude and the very polarity of the resistance response to hydrogenation in high resistivity films. The model of the resistance response that takes into account modifications of thickness was tested and confirmed in three Pd-based systems with variable resistivity: thin Pd films above and below the percolation threshold, thick Pd-SiO$_2$ granular composite films with different content of silica, and Pd-rich CoPd alloys where resistivity depends on Co concentration. Superposition of the bulk resistivity increase due to hydride formation and decrease of film resistance due to thickness expansion provides a consistent explanation of the hydrogenation response in both continuous and discontinuous films with different structures and compositions.

Corresponding author: A. Gerber, email: gerber@tauex.tau.ac.il

**Introduction.**

Volume of the unit cell changes in multiple physical phenomena: due to thermal expansion, under external stress, at phase transitions, due to intercalation and hydrogenation, due to magnetostriction and in many others. When the material is in the form of a thin film grown on a rigid substrate, the lateral in-plane changes are restricted by adhesion to the surface, and the lattice modifications are oriented normal to the plane. This leads to variation of the effective film thickness. Since the changes of the film cross-section are relatively small, their effect on the electrical and thermal conductance properties is usually neglected. We show in the following that modifications of geometry in palladium and palladium based materials absorbing hydrogen can affect significantly the outcome of the experiments and their interpretations, and thus have to be considered.

The catalytic and hydrogen-absorbing properties of palladium make it the metal of choice in almost every aspect of hydrogen-related technologies, including the hydrogen storage and detection. Palladium lattice expands when hydrogen is absorbed, and the volume of free standing material grows as: $\frac{\Delta V}{V} \approx 0.19 \cdot c_H$, where $c_H$ is hydrogen concentration given in H atoms per Pd atom [H/Pd].[1,2] Resistivity of bulk palladium increases in the presence of hydrogen due to the formation of Pd hydride[3-5] having resistivity higher than a pure Pd. Increase of resistivity depends on hydrogen concentration, which is used for constructing the hydrogen metal phase diagrams[6] and in hydrogen detection systems.[7-11] There are, however, multiple reports on the reduction of resistivity at hydrogen loading instead of the anticipated increase. Most of these reports relate to ultrathin[12-15] and nanogranular[16] films and mesowires.[17] Reduction of resistivity is attributed to the discontinuity of the metal structure and expansion of isolated palladium particles following the hydrogenation. It is assumed that in the process of expansion, some particles either touch and form new conducting pathways or reduce the inter-particle gaps, which enhances the inter-particle tunneling and results in an overall decrease in resistance. The necessary condition for implementation of this mechanism is discontinuity of palladium matrix, presence of nanometer void spacers that allow expansion and the very ability of the material to expand laterally within the film plane. This condition is not met in at least a number of cases when a negative resistivity response to hydrogen was observed in solid continuous materials, e.g. in thick Pd films at elevated temperatures[18] or in thick PdCuSi metallic alloys.[16] Convincing explanation of a negative resistivity response to hydrogenation in continuous media has not been suggested so far. We shall demonstrate in the following that the reduction of resistance in both continuous and discontinuous hydrogenated Pd-based films is the result of the expanding film thickness.

**The model.**

Absorption of hydrogen in interstitial sites in bulk free standing Pd yields isotropic lattice expansion:

$$\varepsilon_0 = \frac{\Delta a}{a_0} = \alpha_H c_H,$$

where $a_0$ is the lattice constant of bulk Pd, $c_H$ is hydrogen concentration, $\Delta a$ is the change in the lattice constant $\Delta a = a(c_H) - a_0$, and $\alpha_H$ the expansion factor, $\alpha_H = 0.063$.[1] Films grown on rigid substrates cannot expand laterally within the film plane due to adhesion to the surface. Suppression of the in-plane expansion is equivalent to application of GPa range stresses and the in-plane compression $\varepsilon_{11} = \varepsilon_{22} = -\varepsilon_0$, while the out-of-plane expansion is enhanced by Poisson's effect. Following Wagner et al,[19] the out-of-plane expansion of the [111] textured Pd film is: $\varepsilon_{[111]} \approx 2\varepsilon_0 \approx 2\alpha_H c_H \approx 0.126 c_H$, meaning that thickness of Pd film increases by about 12.6% for $c_H = 1$. Effects of this hydrogenation induced stress and the out-of-plane expansion on microstructure, morphology and deformations of Pd films were studied extensively,[19,20] however the change of thickness was generally neglected in analyzes of the resistance response.[21] The following model takes the modification of geometry into account.

Let's assume a material with the volume absorption of hydrogen and the respective thickness expansion. Resistance of a film before ($R_0$) and after hydrogenation ($R_1$) are given by:

$$R_0 = \frac{\rho_0}{t_0} \cdot \frac{l}{w}$$

and

$$R_1 = \frac{\rho_1}{t_1} \cdot \frac{l}{w}$$

where: $\rho_0$ is the initial resistivity, $t_0$ – the nominal thickness, $l$ and $w$ are the length and width of the film that don't change during hydrogenation because of adhesion to the substrate. Thickness of the hydrogenated film is: $t_1 = t_0 + \Delta t = (1+\gamma)t_0$, where $\gamma$ is the thickness expansion coefficient which depends on the elastic parameters of the given material and the absorbed hydrogen concentration ($\gamma \approx 2\alpha_H c_H$ in [111] textured Pd[19]). Following the Matthiessen's rule, $\rho_1 = \rho_0 + \Delta\rho_H$, where $\Delta\rho_H$ is an additional resistivity scattering term due to absorption of hydrogen and formation of the hydride. For simplicity, we assume here that resistivity is not affected by strain. The change in resistance is:

$$\Delta R = R_1 - R_0 = \frac{\Delta\rho_H - \gamma\rho_0}{(1+\gamma)t_0} \cdot \frac{l}{w} \tag{1}$$

Since the change in resistivity is calculated by using the nominal film thickness $t_0$, it is given by:

$$\Delta\rho = \frac{\Delta\rho_H - \gamma\rho_0}{1+\gamma} \tag{2}$$

Notably, the geometrical (thickness expansion) term depends on the initial resistivity of the material. For thick Pd films with resistivity 10 - 15 µΩcm, depending on the microstructure, and the hydride term $\Delta\rho_H \approx 6$ µΩcm,[22] the negative geometrical contribution is relatively small. However, in high resistivity films, where $\Delta\rho_H \ll \gamma\rho_0$, the geometrical term becomes dominant, the overall resistance response to hydrogen loading is negative, and $\Delta\rho \propto -\rho_0$. Polarity of $\Delta\rho$ is expected to reverse from positive to negative when resistivity exceeds the value: $\rho_{0c} = \Delta\rho_H/\gamma$, which is about 48 µΩcm for [111] Pd at $c_H = 1$.

**Experimental.**

To test the model, we fabricated three different Pd-based systems with variable resistivity: thin Pd films, thick Pd-SiO$_2$ granular films with different content of SiO$_2$, and Pd-rich CoPd alloys where resistivity depends on Co concentration. Series of Pd films with the nominal thickness from 1.2 nm to 100 nm, 100 nm thick Pd-SiO$_2$ films with various content of SiO$_2$, and 15 nm thick CoPd films with different Co concentrations were grown by rf-magnetron sputtering onto room-temperature glass substrates. The base vacuum prior to deposition was $5 \times 10^{-7}$ mbar and Ar pressure during deposition was $5 \times 10^{-3}$ mbar. Binary Pd-SiO$_2$ and CoPd films were co-sputtered from separate targets (2.5" diameter and 2mm thick). The desired composition of the binary samples was controlled by the relative deposition rates in the range 0.01 – 0.1 nm/sec with the respective rf power between 0 and 85 W Pd and CoPd in the deposited films are polycrystalline with prominent fcc (111) texture. Resistance was measured using the Van der Pauw protocol. The hydrogen induced resistance changes were extracted from measurements performed in dry nitrogen and in 4% H$_2$/N$_2$ mixture gas at 1 atm pressure at room temperature. Topology of the selected thin samples was studied by the transmission electron microscopy (TEM) using 5 nm thick films grown directly on TEM grids.

**Results and discussion.**

I. Continuous films.

1) Thin Pd films.

It is well known that the electrical conductivity of very thin metal films fabricated by standard vacuum deposition techniques can be much lower than that of the corresponding bulk material.[23] Direct observations by transmission electron microscopy reveal that such films have a discontinuous structure, with the metal covering only a fraction of the substrate. In the order of decreasing thickness, this structure goes generally through three stages[24]: 1) apparition of isolated holes; 2) coexistence of large holes, metallic islands and a continuous meandric metallic path across the sample; 3) isolated metallic islands. At stage 1, the conductivity is essentially that of the bulk material, modified by diffusive scattering at the surfaces and intergranular boundaries; at stage 2, it can be much lower, though it remains metallic in nature with a positive temperature coefficient of resistivity; at stage 3, the film is an insulator. Conductivity of a continuous meandric metallic structure at stage 2 can be described by the percolation theory[25,26] as:

$$\sigma = \sigma_0(t - t_c)^m \tag{3}$$

where $\sigma_0$ is the bulk conductivity, exponent $m$ is 1.2 for 2-dimensional systems, and $t_c$ is the critical percolation thickness below which metallic clusters become finite. Divergence of resistivity at a finite thickness is the trademark distinguishing the percolation mechanism from the diffusive surface and grain boundaries models.[27-29]

Figure 1 presents the resistivity of a series of Pd films as a function of their thickness (left vertical axis), and the respective change of resistivity in hydrogen environment $\Delta\rho$ (right vertical axis). In films thicker than 50 nm, the resistivity is close to constant with the bulk value $\rho_b$ = 13 µΩcm. Conductivity of films thinner than 10 nm can be well fitted by Eq. 3 with the critical thickness $t_c$ = 2.7 nm and the exponent $m$ = 1.2. The hydrogenation induced change of resistivity $\Delta\rho$ is defined as:

$$\Delta\rho = \Delta R \cdot tw/l = (R_{H_2} - R_{N_2}) \cdot tw/l,$$

where $R_{N_2}$ and $R_{H_2}$ are the resistance values measured in a pure nitrogen atmosphere and in nitrogen/hydrogen mixture with 4% of $H_2$, respectively. $t$ is the film nominal thickness. As seen, the data follow the predictions of Eq. 2. $\Delta\rho$ is positive in low resistivity thick films and becomes negative in thin high resistivity films. The highest positive value, $\Delta\rho$ = 6 µΩcm is found in the thickest 100 nm film in agreement with Ref. 22. The cross-over from positive to negative response occurs at about 7 nm thickness for which the resistivity is about 42 µΩcm. Topology of the 7 nm thick film (TEM not shown) corresponds to the stage 2 mentioned above: there are isolated voids in the otherwise continuous metallic film. It is important to note that all samples presented in Fig.1 demonstrate positive resistivity temperature coefficients $d\rho/dT > 0$, meaning that conductance occurs along continuous metallic media. Similar crossover from positive to negative resistivity response in thin Pd films has been reported in Refs [30,31]. There, the effect was interpreted as due to transition between the continuous and discontinuous film topology, while the negative resistivity response in the discontinuous case was explained by the in-plane lateral swelling of Pd clusters.[30-32]

2) Pd-SiO$_2$ granular mixtures.

To resolve the question whether the in-plane or the out-of-plane swelling is the decisive mechanism of the hydrogenation response, we used a different system where resistivity of thick films can be modulated without opening voids in the material volume. The material is Pd co-deposited with different amounts of insulating SiO$_2$. SiO$_2$ is immiscible in Pd, therefore Pd-SiO$_2$ is a granular mixture of nanocrystalline metallic Pd and amorphous insulating SiO$_2$. Transmission electron microscope (TEM) micrograph of 5 nm thick Pd$_{70}$ - (SiO$_2$)$_{30}$ film with 70% vol. of Pd and 30% vol. of SiO$_2$ is shown in Fig.2. Polycrystalline Pd (dark in the figure) with crystalline dimensions of 3 – 5 nm form a typical percolation pattern with amorphous SiO$_2$ filling the voids. Fig.3 presents the resistivity of a series of 100 nm thick Pd-SiO$_2$ films (right vertical axis) and the normalized resistivity response to hydrogenation (left vertical axis) as a function of SiO$_2$ volume concentration $x$. Resistivity of Pd-SiO$_2$ increases with silica concentration and exceeds 500 μΩcm for x = 30%. Hydrogenation induced change of resistivity is positive in pure Pd and changes gradually to negative with increasing SiO$_2$ content in agreement with Eq.2. Transition from positive to negative response occurs at SiO$_2$ concentration of about 17% when resistivity exceeds the 64 μΩcm mark. All samples shown here demonstrate positive resistivity temperature coefficients $d\rho/dT > 0$ and are above the metallic continuity percolation threshold. Silica fills the voids between palladium, therefore the in-plane swelling of individual Pd clusters through SiO$_2$ can be excluded in this case. Development of a negative response to hydrogenation is attributed solely to the expansion of thickness.

3) CoPd alloys.

Equation 2 predicts that the resistivity response of any expanding in hydrogen material is negative if its initial resistivity is sufficiently high. To test the generality of this prediction, we examined one more system: the CoPd alloys. Ferromagnetic CoPd alloys attracted attention recently[33,34] as the material for magnetic detection of hydrogen using the extraordinary Hall effect. Resistivity of the alloy films depends on Co content, as illustrated in Fig.4 for a series of 15 nm thick samples. Starting from a pure Pd film, resistivity increases to 73 µΩcm at Co atomic concentration x = 47%. The respective resistivity response to hydrogenation is shown by solid circles in Fig.4 (left vertical axis). Similar to the two previous cases, the resistivity response reverses from positive in pure Pd to negative in high resistivity CoPd alloys when Co concentration exceeds 13%.

Fig.5 summarizes the resistivity response to hydrogenation of thin continuous Pd films, thick continuous Pd-SiO$_2$ granular mixtures and CoPd alloys as a function of the material resistivity. All systems behave in the same way: the response to hydrogenation varies gradually from positive in low resistivity thick Pd and Pd-rich films to negative in high resistivity samples regardless of their structure and composition. The cross-over resistivity is 42 µΩcm in percolating Pd films, 64 µΩcm in Pd-SiO$_2$ and 38 µΩcm in CoPd. These values are remarkably close to each other and to the estimation of 48 µΩcm made for the textured Pd at $c_H = 1$. It is important to note that the exact value of $c_H$ is not known in these experiments. One can conclude that although the thickness expansion is relatively small, its effect on the resistance response to hydrogenation is dominant in high resistivity materials.

II) Continuous versus discontinuous films.

An obvious question to ask is whether the mechanism of a negative resistivity response to hydrogenation in solid continuous films is different from the discontinuous ones, in other words whether the in-plane or the out-of-plane expansion is the mechanism dominant in thin discontinuous films. The lateral in-plane expansion of hydrogenated Pd is prohibited at the interface with the substrate but can occur when the distance from the substrate grows. Finite element simulation of the volume expansion of the initially flat Pd islands predicted considerable in-plane expansion, growing with the island height.[35] A gap of about 14 nm width can be bridged between adjacent islands if they are 30 nm thick. In 2 nm thick films the lateral expansion reduces to negligibly small.

To clarify the point, we extended the measurements to disconnected Pd structures below the percolation threshold in ultrathin films and in thick Pd-SiO$_2$ mixtures with high SiO$_2$ concentration. Measurements of resistivity as a function of temperature were used to establish the continuity of metallic films. Excluding the low temperature range, resistivity of metals increases with temperature with a positive resistivity temperature coefficient $d\rho/dT > 0$. When metallic film is discontinuous, insulating gaps appear across the conductance paths. If thickness of these gaps is small the charge is transferred by thermally assisted tunneling, characterized by a negative resistivity temperature coefficient $d\rho/dT < 0$. Since the total resistance of the conducting network is determined by its highest value components, i.e. the tunnel junctions if present, polarity of the resistivity temperature coefficient indicates the continuity or discontinuity of the metallic structure. Thus, $d\rho/dT > 0$ indicates a continuous metallic structure, while $d\rho/dT < 0$ the discontinuous one.

Typical temperature dependence of the normalized resistivity as a function of temperature between the room and 77K is shown in Fig. 6 for several thin Pd (Fig. 6a) and thick Pd-SiO$_2$ (Fig. 6b) films. The 3 nm thick Pd film (Fig.6a) demonstrates a positive resistivity temperature coefficient $d\rho/dT > 0$ in the entire tested temperature range down to 77 K and has, therefore a continuous Pd structure. The 1.5 nm and 1.9 nm thick films are discontinuous with $d\rho/dT < 0$. The 2.4 nm thick film demonstrates a positive $d\rho/dT > 0$ at room temperature, the resistivity minimum at 130 K (not seen at the figure resolution) and negative $d\rho/dT < 0$ at lower temperatures. The resistivity minimum in granular composite and thin films was identified[36] as the onset of the tunneling dominated regime, below which the temperature dependent intragranular conductance exceeds the temperature dependent intergranular tunneling conductance, and the system enters the weakly insulating regime. A positive resistivity temperature coefficient above the minimum indicates the intragranular metallicity when intragranular conductance is smaller than the intergranular tunneling and the intragranular resistivity dominates the total. Following this interpretation, granular films demonstrating a resistivity minimum are below the geometrical percolation threshold. Thus, the entire series represent the samples above, at and below the continuity threshold. Similarly, in thick granular $Pd_{100-x}$ - $(SiO_2)_x$ films (Fig. 6b), Pd matrix is continuous above the percolation threshold in samples x = 30% and x = 40%, and discontinuous in samples x = 50% and x = 60%. Fig. 7 presents the reduction of resistivity in hydrogenated samples as a function of resistivity for the entire range of continuous and discontinuous thin Pd and thick Pd-SiO$_2$ films. Starting from the resistivity about 1 mΩcm and for more than four orders of magnitude beyond 10 Ωcm, the negative response to hydrogenation is proportional to the initial resistivity, in agreement with Eq. 2. The discontinuous samples follow the same dependence as the continuous ones. Below the percolation, the measured resistance is due to the intergranular tunneling. The cross section of tunnel junctions increases with the expanding Pd thickness at the same ratio as the cross section of continuous films, therefore the correlation between the hydrogenation response and the initial resistivity remains constant in both continuous and discontinuous films. Obviously, one cannot exclude random changes in the in-plane profile of discontinuous ultrathin Pd films. However, a systematic correlation between the hydrogenation response and resistivity over more than four orders of magnitude indicates that the same mechanism is responsible for the reduction of resistivity in both continuous and discontinuous Pd matrices.

**Summary.**

To summarize, we developed a simple model of the hydrogenation induced changes of resistance in Pd based materials, which takes into account an expansion of the effective film thickness under the lateral in-plane stress. The change of resistance is predicted to depend on the initial resistivity of the material and to reverse its polarity from positive (increase of resistance in hydrogenated material) in low resistivity samples to negative (decrease of resistance) in high resistivity films. The model was tested in three Pd-based systems with variable resistivity: thin Pd films, thick Pd-SiO$_2$ granular composite films with different content of silica, and Pd-rich CoPd alloys where resistivity depends on Co concentration. All systems demonstrated a similar gradual transition from a positive response to hydrogenation in low resistivity samples to a negative one when their resistivity exceeded 40 - 60 μΩcm, very close to the model estimation. Reduction of resistivity in the hydrogenated state is proportional to the initial resistivity of high resistivity samples in both continuous and discontinuous Pd matrices, which indicates that the out- of-plane and not the lateral in-plane Pd expansion is the source of the negative resistance response in thin films. Thus, superposition of the bulk resistivity increase due to hydride formation and decrease of film resistance due to thickness expansion provides a consistent explanation of the hydrogenation response in both continuous and discontinuous films with different structures and compositions.

## Acknowledgements.

The research was supported by the State of Israel Ministry of Science, Technology and Space grant No.3217522.

## Data availability.

The data that support the findings of this study are available from the corresponding author upon reasonable request.

# Figure captions

Fig.1. Resistivity (open circles, left vertical axis) and the respective change of resistivity in hydrogen environment (solid circles, right vertical axis) of a series of Pd films as a function of their thickness. Symbols size is larger than the error bars in all figures.

Fig.2. Transmission electron microscope (TEM) micrograph of 5 nm thick $Pd_{70}$ - $(SiO_2)_{30}$ film with 70% vol. of Pd and 30% vol of $SiO_2$ . Polycrystalline Pd (dark in the figure) with crystalline dimensions of 3 – 5 nm form a typical percolation pattern with amorphous transparent $SiO_2$ filling the voids.

Fig. 3. Resistivity of 100 nm thick Pd-$SiO_2$ films (open circles, right vertical axis) and the normalized resistivity response to hydrogenation (solid circles, left vertical axis) as a function of $SiO_2$ volume concentration $x$.

Fig. 4. Resistivity of 15 nm thick CoPd alloy films (open circles, right vertical axis) and the respective resistivity response to hydrogenation (solid circles, left vertical axis) as a function of Co atomic concentration $x$.

Fig. 5. The normalized resistivity response to hydrogenation of thin continuous Pd films, thick continuous Pd-$SiO_2$ granular mixtures and CoPd alloys as a function of the samples resistivity.

Fig. 6. Normalized resistivity as a function of temperature in (a) thin Pd films of different thickness; and (b) 100 nm thick Pd-$SiO_2$ films with different $SiO_2$ volume concentration $x$.

Fig.7. Reduction of resistivity $-\Delta\rho$ in hydrogenated samples as a function of resistivity for continuous and discontinuous thin Pd and thick Pd-SiO$_2$ films. Open symbols indicate the continuous Pd matrices with positive resistivity temperature coefficients. Solid symbols relate to discontinuous Pd structures with the temperature assisted tunneling and negative resistivity temperature coefficients. The solid line is the guide for eyes.

**Figures.**

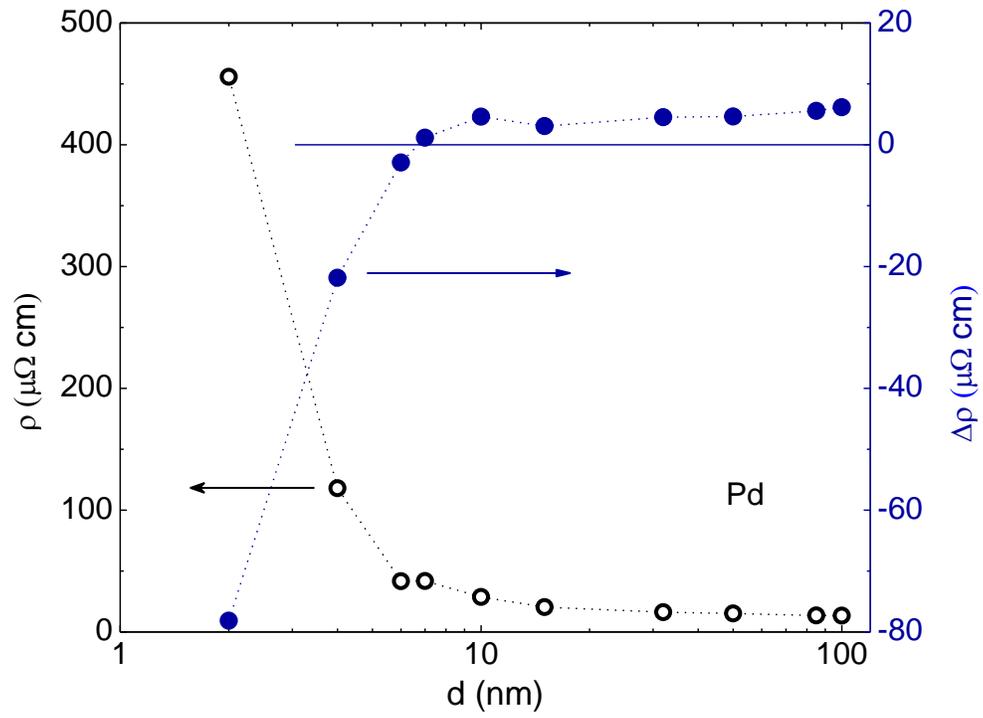

Fig. 1

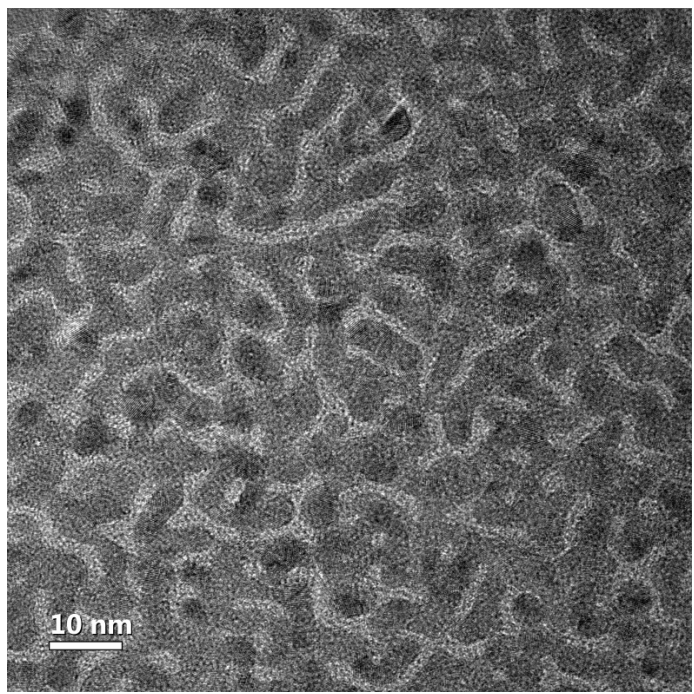

Fig. 2

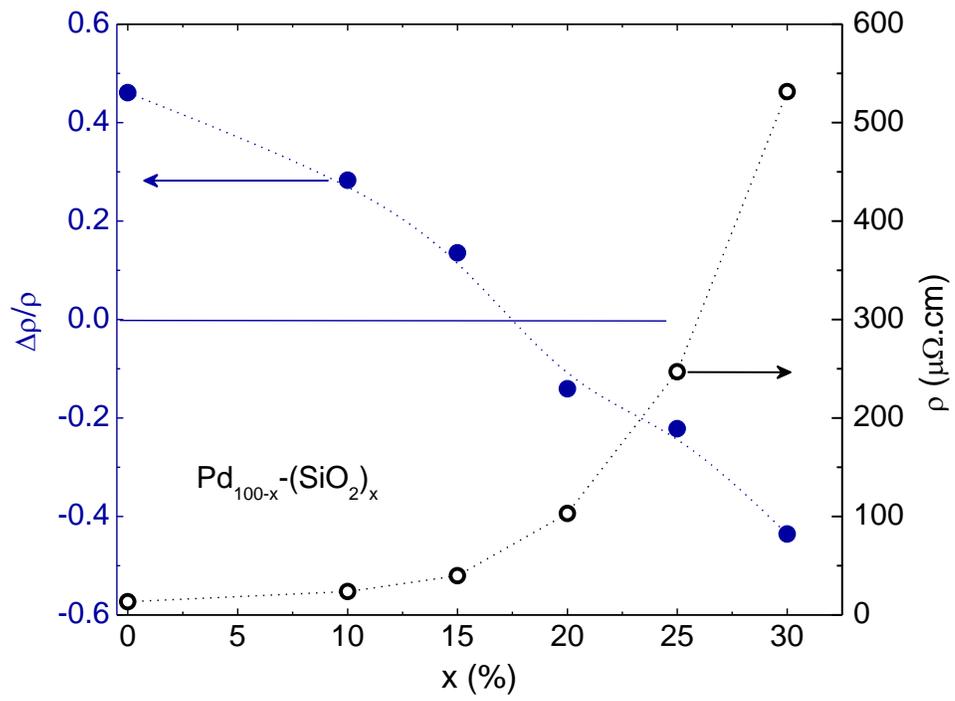

Fig.3

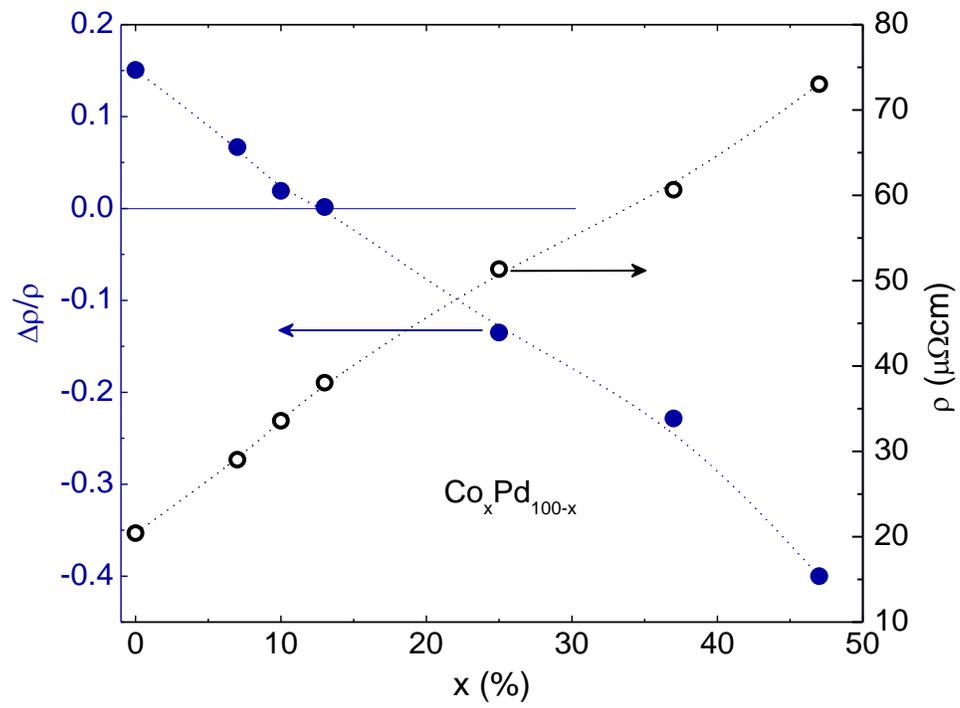

Fig. 4

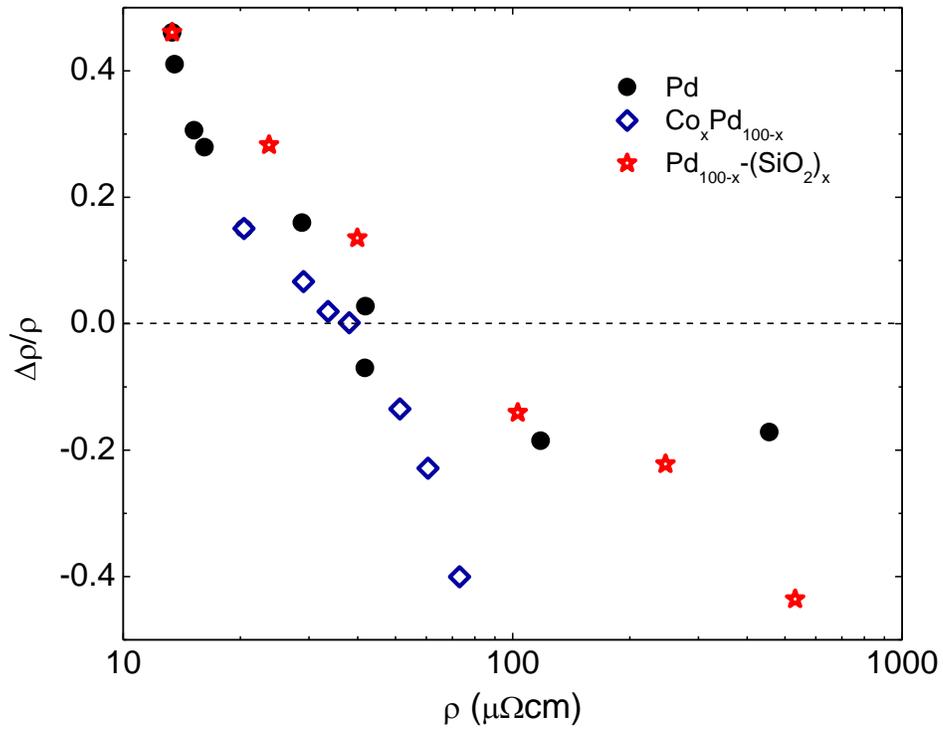

Fig. 5

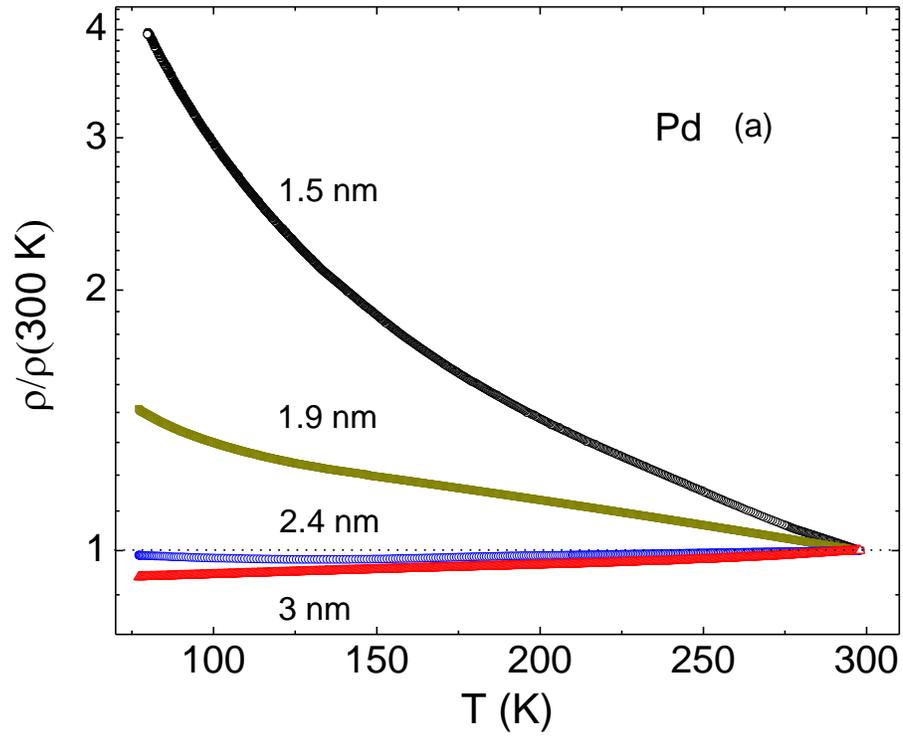

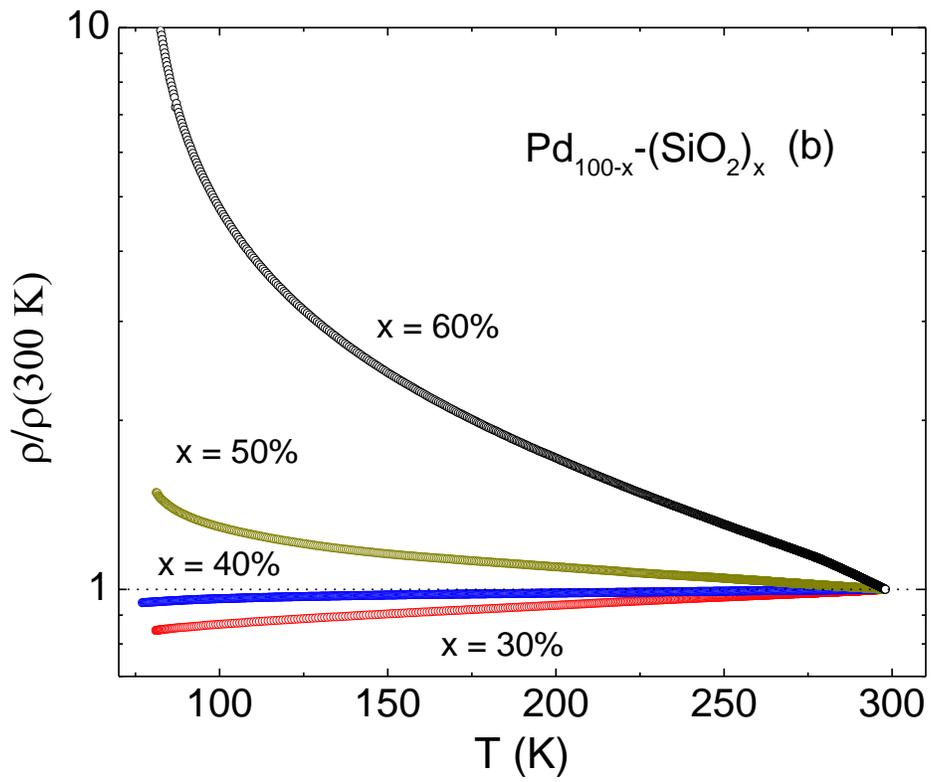

Fig. 6

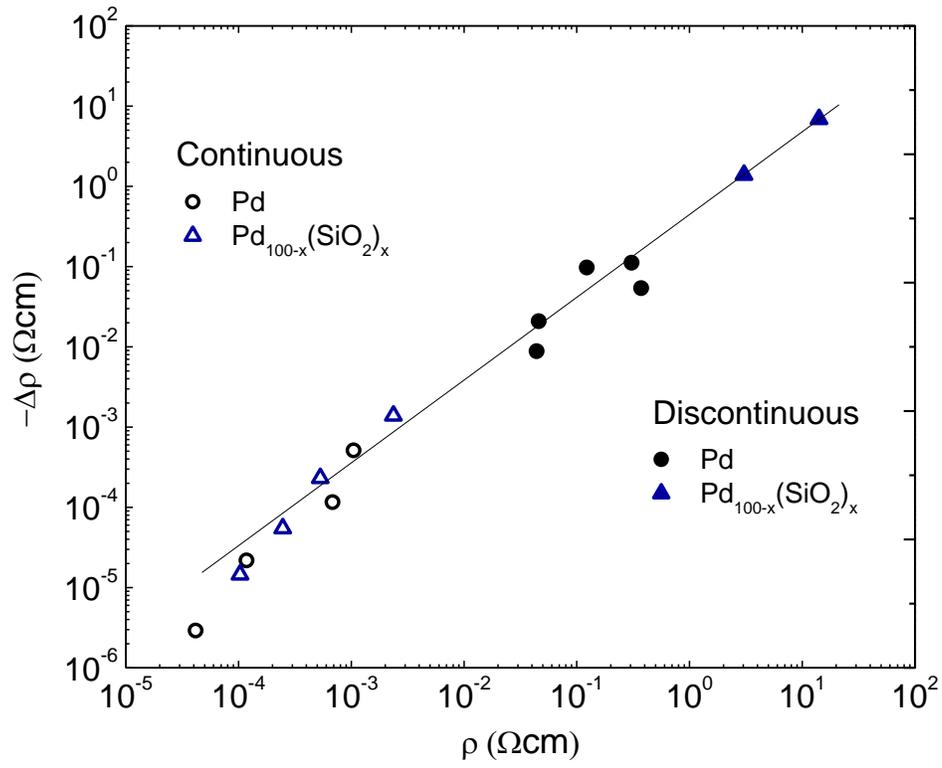

Fig. 7